\documentstyle[12pt,epsf]{article}
\input{psfig.tex}
\begin{document}
\begin{titlepage}
\Large
\center{POLARIZATION OF THE MICROWAVE BACKGROUND IN REIONIZED MODELS}
\vspace{0.5cm}
\normalsize
\center{Matias Zaldarriaga\footnote{Email address:
matiasz@arcturus.mit.edu}} 
\vspace{0.3cm}
\center{
Department of Physics, MIT, Cambridge, MA 02139 USA}
\vspace{0.3cm}
\begin{abstract}
I discuss the physics of polarization in
models with early reionization. For sufficiently high optical
depth to recombination
the polarization is boosted on large scales
while it is suppressed on smaller scales. New peaks appear in the 
polarization power spectrum, their position is proportional
to the square root of the redshift at which  the reionization
occurs while their amplitude is proportional to the optical depth. 
For standard scenarios the $rms$ degree of linear polarization as
measured with a  $7^o$ FWHM antenna  (like the one of the 
Brown University experiment) is $1.6\mu K$ , $1.2 \mu K$ ,
$4.8\times 10^{-2} \mu K$ 
for an optical depth of $1$, $0.5$ or $0$ respectively.
For a $1^o$ FWHM antenna this same models give $2.7 \mu K$ , $1.8 \mu K$ 
and $0.77 \mu K$. Detailed measurement of polarization on large
angular scales could provide an accurate determination of the
epoch of reionization, which cannot be obtained from temperature
measurements alone.
\end{abstract}
\end{titlepage}
\newpage
\section{Introduction}

Since the first measurements of cosmic microwave background (CMB)
anisotropies by the COBE satellite a few years ago this field has seen
a very rapid development. There have been a number of new 
detections on smaller angular scales
\cite{bond96,scott95}
as well as a lot of progress on the theoretical side 
\cite{bond94,hu95c,seljak94}. 
Proposed microwave background experiments may be able to measure 
cosmological parameters with great accuracy, although some of
the parameters may be degenerate \cite{bond94}.

The polarization of the microwave background has also received 
attention. On the theoretical side
the polarization induced by density perturbations in models
with a standard ionization history has been studied both numerically
\cite{be87} and analytically \cite{zal95}.
The possibility of using polarization 
to distinguish between scalar and tensor
fluctuations has been investigated \cite{crittenden93,ng95,har93}. 
The temperature-polarization cross
correlation function for tensor modes has also been studied as a possible
probe of the importance of the tensor contribution to the CMB anisotropies
\cite{crittenden94}. More recently the possibility of 
using the CMBR polarization
to measure primordial magnetic fields has been investigated
\cite{loeb96}. 

On the experimental 
side, there have been a number of experiments 
\cite{penzias65,nanos79,caderni78,lubin81}.
An upper limit of $6\times 10^{-5} \mu K$
on the degree of linear polarization has been established. 
An experiment to measure CMB polarization 
is now under construction at Brown University.
\footnote[1]{ Visit
http://www.physics.brown.edu/ObsCosmology/polarize/polarize.html 
for details.}
It will measure the $Q$ and $U$ Stokes parameters using first a 
$7^o$ FWHM antenna and then a $1^o$ one. The expected sensitivity
of this instrument is of a few $\mu K$. The future satellite missions
MAP and COBRAS/SAMBAS will also measure polarization 
\cite{map,cosa}.

It was soon realized that an early reionization of the universe
will greatly enhance polarization \cite{be84}. The fact
that in universes that never recombined the polarization would also be
large was noted in many of the above studies. More recently 
Ng \& Ng \cite{ng95} discussed the polarization generated
in reionized universes with instantaneous  recombination. 
The Sachs-Wolfe effect was the only source of anisotropies that they
included. They
concluded that reionization at a moderate redshift could boost polarization
to the level of a few percent of the temperature perturbations.
Although this conclusion is correct, to make detailed predictions
for an experiment such as that being built at Brown a realistic
recombination history should be used since polarization is very 
sensitive to the duration of recombination \cite{zal95,frewin94}. 
Baryons should also be included in the 
calculation as the acoustic oscillation in the photon-baryon
plasma are very important to determine polarization.   

In this paper I discuss in detail the physics behind the
polarization generated in models where there was an early 
reionization after the usual recombination. 
These models
show very distinct features in the polarization power spectrum 
including 
a new peak at low $l$. This
peak is not present either in the standard recombination scenarios or 
in the cases where the universe never recombined and it is the cause of the
boost in the polarization.

All the calculations where done using the code CMBFAST
\footnote[2]{This code is publically available, for a copy contact
Uro\v s Seljak (useljak@cfa.harvard.edu) or the author.} recently developed
by Uro\v s Seljak and the author \cite{seljak96a}. This code 
is both fast and accurate so detailed predictions for the Brown
experiment or future satellite missions like MAP
can be easily obtained.   
 
\section{Standard Ionization History}

In this section I review previous results for the CMB polarization
for a standard ionization history in a flat space-time. 

The anisotropy and polarization perturbations can be expanded in 
terms of Fourier modes, which are independent in the linear regime.
For one mode with wavevector $\vec k$ $\Delta_T(\vec k, \vec n)$ and
$\Delta_P(\vec k, \vec n)$ will denote the temperature and polarization
perturbations, where  $\vec n$
is the direction of photon propagation.  
The perturbations can be further expanded in Legendre series,
\begin{equation}
\Delta(\vec k, \vec n)=\sum_l (2l+1)(-i)^l\Delta_{l} P_l(\mu),
\label{delta}
\end{equation} 
where $\mu=\vec k \cdot \vec n/k$. This expansion applies both to the 
anisotropy and polarization perturbation \cite{be84,crittenden93,kos96}. 

The Boltzmann equations for the perturbations in the scalar case 
are given by 
\cite{be87},
\begin{eqnarray} 
\dot\Delta_T +ik\mu \Delta_T 
&=&\dot\phi-ik\mu \psi+\dot\kappa\{-\Delta_T +
\Delta_{T0} +i\mu v_b +{1\over 2}P_2(\mu)\Pi
\} \nonumber \\   
\dot\Delta_P +ik\mu \Delta_P &=& \dot\kappa \{-\Delta_P +
{1\over 2} [1-P_2(\mu)] \Pi\} \nonumber \\
\Pi&=&\Delta_{T2}+\Delta_{P2}+\Delta_{P0}.
\label{Boltzmann}
\end{eqnarray}
Here the derivatives are taken with respect to the conformal time $\tau$ 
and $v_b$ is the velocity of baryons. The 
differential optical depth for Thomson scattering is denoted as 
$\dot{\kappa}=an_ex_e\sigma_T$, where $a(\tau)$ 
is the expansion factor normalized
to unity today, $n_e$ is the electron density, $x_e$ is the ionization 
fraction and $\sigma_T$ is the Thomson cross section. The total optical 
depth at time $\tau$ is obtained by integrating $\dot{\kappa}$,
$\kappa(\tau)=\int_\tau^{\tau_0}\dot{\kappa}(\tau) d\tau$.
A useful variable is the visibility function $g(\tau)=\dot{\kappa}
{\rm exp}(-\kappa)$. For a standard ionization history it's peak  
defines the epoch of decoupling, when the  
dominant contribution to the CMB anisotropies arises.

This equations can be formally integrated to give
(\cite{seljak96a}  and references therein),
\begin{eqnarray}
\Delta_T &=&\int_0^{\tau_0}d\tau e^{i k \mu (\tau -\tau_0)}
e^{-\kappa} 
\{ \dot\kappa [\Delta_{T0}+i\mu v_b + {1\over 2} P_2(\mu)
\Pi]+\dot\phi-ik\mu\psi\} \nonumber \\	
\Delta_P &=& -{1\over 2}\int_0^{\tau_0} d\tau e^{i k \mu (\tau -\tau_0)}
e^{-\kappa} \dot\kappa   [1-P_2(\mu)]
\Pi .
\label{formal}
\end{eqnarray}
Equation (\ref{formal}) is the basis for the line of sight approach used 
in CMBFAST.

The temperature anisotropy 
spectra, $C_{Tl}$ 
is  defined as 
\begin{equation}
C_{Tl}=(4\pi)^2\int k^2dk P_\psi(k)|\Delta_{Tl}(k,\tau=\tau_0)|^2.
\label{cl}
\end{equation}
where
$P_\psi(k)$ is the power spectrum of the metric perturbations.

The temperature angular correlation function is related to the 
temperature $C_{Tl}$ power spectrum by
\begin{equation}
C(\theta)=\langle \Delta T(\vec n_1)\Delta T (\vec n_2)\rangle_{\vec n_1
\cdot
\vec n_2=\cos \theta}={1 \over 4\pi}\sum_{l=0}^\infty(2l+1)C_{Tl}
P_l(\cos \theta).
\label{ctheta}
\end{equation}

Because the polarization in a tensor quantity the expressions 
for the correlation functions are somewhat more complicated.
Polarization can be analized using spin-weighted spherical
harmonics \cite{zalsel}, when considering the polarization produced
by density perturbations only one power
spectra, $C_{El}$, is enough to characterize polarization statistics,
\begin{eqnarray}
C_{El}^{(S)}&=&
(4\pi)^2\int k^2dkP_\psi(k)\Big[\Delta_{El}(k)\Big]^2
\nonumber \\
\Delta_{El}(k)&=&\sqrt{(l+2)! \over (l-2)!}\int_0^{\tau_0} d\tau 
S_{E}(k,\tau) j_l(x) \nonumber \\  
S_E(k\tau)&=& 
{3g(\tau)\Pi(\tau,k) \over 4 x^2},
\label{es}
\end{eqnarray}
where  $j_l$ denote the spherical Besssel functions and $x=k(\tau_0-\tau)$. 

The root mean square fluctuations are given by
\begin{equation}
\langle P^2 \rangle\equiv\langle (Q^2+U^2) \rangle=2\langle Q^2 \rangle=
{1 \over 4\pi}
\sum_{l=0}^\infty(2l+1)C_{El} W_{l}.
\end{equation} 
$P$ is just the degree of linear polarization and $W_{l}$ is the 
window function for the particular experiment under consideration.
 
Figure 1 shows the temperature and polarization 
$C_l$ spectra obtained by numerically integrating
the above equations using CMBFAST, 
for the standard CDM model ($\Omega_0=1$, $H_0=50\, \rm km\,\rm  sec^{-1}$
and $\Omega_{b}=0.05$), normalizing the result to the COBE measurement.
Normalization was carried out using the fits to the shape 
and amplitude of the 4 year COBE data described in \cite{bunnwhite}, this
aproximately  fixes $10\times 11 \times C_{T10}/2\pi\sim 830 \mu K^2$. 

The features in the polarization power spectrum can be understood 
analytically \cite{zal95}. 
Polarization is produced by Thomson 
scattering of anisotropic radiation. To be more precise, the source of 
polarization is the quadrupole of the intensity distribution in the
rest frame of the electrons, $\Delta_{T2}$ in equation (\ref{Boltzmann}).
Thus no polarization can be generated after decoupling if there 
is no reionization or anisotropy. Before recombination 
the photons and baryons were tightly coupled,
the damping scale being
only a few Mpc. For this reason the photon
distribution function was nearly isotropic in the rest frame of the 
electrons and thus the generated polarization was extremely small. 
As photons and electrons decouple, the mean free path of the photons
starts to grow and temperature quadrupole moment is produced
by free streaming.
Now photons scattering off a given electron come from 
regions where electrons have slightly different velocities, {\it i.e.} 
the redshift of these photons and thus the intensity at a fixed wavelength
depends on direction. The quadrupolar part of this temperature 
fluctuations is the source of the generated polarization.
For wavelengths longer than the width of the last scattering surface,
$\Delta\tau_D$,
the polarization perturbation can be shown to be
\cite{zal95}, 
\begin{equation}
\Delta_P =0.51  
(1-\mu^2) e^{i k \mu (\tau_D -\tau_0)} k\Delta\tau_D \Delta_{T1}(\tau_D) 
\label{analitzh}
\end{equation}
$\tau_D$ is the conformal time of decoupling.
Note that in the tight coupling regime $\Delta_{T1}\propto v_b$.
The above formula shows that for wavelengths longer than the width of
the last scattering surface, 
polarization is proportional to the velocity 
difference between places separated by a distance $\Delta\tau_D$, 
the distance photons travel on average during decoupling. 

For the standard adiabatic initial conditions 
$\Delta_{T1}$ and the baryon velocity  vanish as $k \tau \rightarrow 0$
which together with the $k\Delta\tau_D$ factor in the previous expression
explain the dramatic fall of polarization for large angular scales. 
For large wavelengths the quadrupole generated in the photon 
distribution as photons travel
between their last scatterings is extremely small  both due to the small
distance they can travel compared to
the wavelength as well as  to the small velocity
differences generated by these small $\vec k$ perturbations. 

For smaller angular scales, $l \ge 100$,
the same acoustic oscillations that
generate the Doppler peaks in the temperature anisotropy cause
the peaks in the polarization spectrum. The peaks are located at 
different $l$ values because they occur for different wavevectors.
The anisotropy peaks correspond to the maxima of the
temperature monopole \cite{hu95a,hu95b,hu95c}
while 
from (\ref{analitzh}) those in the polarization occur 
at the maxima of  the temperature dipole, {\it i.e.} the baryon velocity. 
In the tightly coupled regime,
the temperature dipole is proportional to the time derivative of the 
monopole which explains the fact that polarization peaks
occur at the $l$ values where the temperature is at is minima.

For smaller scales Silk 
damping damps the oscillations in the photon baryon plasma and this
together with cancellations due to the finite width of the last 
scattering surface,
is the cause for the decay in the $C_l$ spectrum for both temperature and 
polarization (Figure 1). 

\section{The Reionized Case}

In this section I  consider models with early reionization
and try to explain the origin 
of the new features that appear in the polarization power spectrum.

For definitiveness I  use a  
standard CDM model
where the universe reionized at an epoch such that the optical depth to 
recombination is $\kappa_{ri}$. This means for example that reionization 
occurred at a redshift of around $z_{ri}\sim100$ if $\kappa_{ri}=1.0$. 
Figure 2 shows the visibility function, $g(\tau)=\dot\kappa \exp{(-\kappa)}$, 
for $\kappa_{ri}=1.0$ assuming that all hydrogen atoms are ionized
up to the present epoch ($x_e=1.0$). The visibility
function has a very simple interpretation, the probability that a photon 
reaching the observer last scattered between $\tau$ and $\tau + d\tau$
is just $g(\tau) d\tau$. The first peak in figure 2, occurring at 
$\tau \approx 120 Mpc$ for sCDM ($h=0.5$)
accounts for the photons that last scattered
at recombination, the area under this peak, the probability
that a photon came directly to us from this epoch, is $\exp(-\kappa_{ri})$.
The area under the second peak gives the fraction of photons 
that scattered after reionization before reaching the 
observer, and is equal to $1-\exp(-\kappa_{ri})$. 

Figure 1 shows the result of numerically integrating the Boltzmann
equations using CMBFAST for this reionized case. On small angular scales,
the polarization  ``Doppler peaks'' are suppressed, just as those in 
the anisotropy are. This is very simple to understand, only
a fraction $\exp(-\kappa_{ri})$ of the photons reaching the observer come from
recombination, so their contribution to the $C_l$ power spectrum is reduced 
by a factor $\exp(-2 \kappa_{ri})$. 
On large angular scales new peaks appear in the polarization 
power spectrum. The temperature anisotropy shows 
no new peaks. This peaks are what boost the polarization
on large scales and may take it to detectable levels.

Let us  try to understand the origin of these peaks. For 
low values of $k$ the 
largest perturbation in the photon distribution function is the monopole,
$\Delta_{T0}$
because of 
 the tight coupling between photons and electrons before recombination.
Both the dipole and the quadrupole as well as the polarization
perturbations are much smaller. But after photons and electrons decouple, all
the temperature multipoles can grow by free streaming. Power is being
carried from the zero multipole 
moment to higher ones, which is just a geometrical 
effect. The
temperature quadrupole is growing by free streaming after recombination and 
so by the time of reionization there is and appreciable quadrupole that can
generate polarization. The structure of this quadrupole explains the new 
features in the polarization power spectrum.

The formal line of
sight solution for the polarization perturbation is
\begin{equation}
\Delta_P=-{1\over 2}\int_0^{\tau_0} d\tau e^{i k \mu (\tau -\tau_0)}
e^{-\kappa} \dot\kappa   [1-P_2(\mu)]
\Pi .
\label{lospol}
\end{equation}
The visibility function, $\dot\kappa e^{-\kappa}$, has two peaks one
at recombination and the other due to reionization, so it is convenient
to separate the previous integral in two parts,
\begin{equation}
\Delta_P=-{1\over 2}[1-P_2(\mu)]( \int_0^{\tau_{ri}} 
d\tau e^{i k \mu (\tau -\tau_0)} \dot\kappa e^{-\kappa}  \Pi +
\int_{\tau_{ri}}^{\tau_{0}} 
d\tau e^{i k \mu (\tau -\tau_0)} \dot\kappa e^{-\kappa} 
\Pi) 
\label{polsep}
\end{equation}
where $\tau_{ri}$ is the conformal time of the start of reionization.
The first integral just represents the polarization generated at 
recombination and can easily be shown to be 
\begin{equation}
\Delta_P^{(1)}\equiv-{1\over 2}[1-P_2(\mu)] \int_0^{\tau_{ri}} 
d\tau e^{i k \mu (\tau -\tau_0)} \dot\kappa e^{-\kappa} \Pi =
e^{-\kappa_{ri}}\Delta_P^{NR} 
\end{equation}
where $\Delta_P^{NR}$ is the polarization that would be measured
if there was no reionization, as  discussed in the previous section.
This contribution is damped because only a fraction $\exp(-\kappa_{ri})$
of the photons that arrive to the observer came directly from recombination
without scattering again after reionization.

Let us now consider the new contribution arising from reionization.
The polarization source is $\Pi=\Delta_{T2}+\Delta_{P2}+\Delta_{P0}$. 
$\Delta_{T2}$ is large coming  from the free 
streaming of the monopole at recombination, while the
polarization terms do not grow after decoupling 
and are thus negligible to  first approximation. Equation (\ref{polsep})
shows that the new polarization is basically an average of the value
of the temperature quadrupole during the reionization scattering surface.
This accounts for all the new features in the polarization power spectrum
of Figure 1.

To understand the origin of these new  peaks
let us find the amplitude of the temperature quadrupole at the time 
reionization starts $\tau_{ri}$. The monopole at recombination 
is approximately given by \cite{hu95c}
\begin{equation}
(\Delta_{T0}+\psi)(\tau_D)={1\over 3}\psi (1+3R) \cos(k c_s \tau_D) -R \psi 
\end{equation}
$\psi$ is just the value of the gravitational potential (assumed 
constant), $R=3\rho_b / 4 \rho_{\gamma}|_{\tau_D}
\approx 30 \Omega_b h^2$ and 
$c_s=1 / \sqrt{(1+R)}$ is the photon-baryon sound speed.
The quadrupole at $\tau_{ri}$ arising from the free streaming 
of this monopole is simply
\begin{equation}
\Delta_{T2}(\tau_{ri})= (\Delta_{T0}+\psi)(\tau_D) j_2 [k(\tau_{ri}-
\tau_D)]
\label{dt2}
\end{equation}
where $j_2$ is the $l=2$ spherical Bessel function. 

The peaks of the previous expression as a function of $k$
will show up in the polarization power spectrum. The first 
peak of (\ref{dt2}) is approximately at the first peak of
the Bessel function because $ c_s \tau_D \ll (\tau_{ri}-\tau_D)$.
The wavevector for this 
first peak is approximately given by
$k(\tau_{ri}-\tau_D)\sim 2$,
these wavevector translates into an  $l$ value as usual according to 
$l\sim k (\tau_0-\tau_{ri})$ and thus the $l$ value for the
first is $l \sim 2 (\tau_0-\tau_{ri}) /
(\tau_{ri}-\tau_D)\sim 2  \sqrt{z_{ri}}$.
For the case under consideration this means 
$l\sim 24$ which agrees very well with the 
the first peak in Figure 3. Only the first peaks appear because
the reionization scattering surface is very wide and thus 
the integrand
in equation (\ref{polsep})
for smaller wavelengths oscillates during its width and cancels out after
integration. 
This cancellation
makes the new polarization small and thus  hidden under the polarization
generated at recombination.
 
The major factor determining the difference in height 
of these new peaks for different models  
is the fraction of photons reaching the observer 
that last scattered after reionization, $1-\exp(-\kappa_{ri})$.
Thus the ratio of the distances
between the observer and reionization to that between the two scattering 
surfaces determines the positions of the peaks, and the optical depth 
$\kappa_{ri}$ their heights.
 
To further illustrate these points Figure 3a show the $C_{El}$ spectrum for 
standard CDM models with varying optical depths $\kappa_{ri}$. The peaks not 
only vary in height but also in position, as the redshift of reionization 
has to increase in order to increase $\kappa_{ri}$, thus the ratio of 
distances that determines the position of the peaks gets bigger, 
as  $(\tau_0-\tau_{ri})$ increases and $(\tau_{ri}-\tau_D)$ decreases,
driving the peaks to a smaller angle ($l_{peak}\propto \sqrt{z_{ri}}$). 

Figure 3b on the other hand show how these peaks vary with the cosmological 
constant for a fixed reionization redshift $z_{ri}=100$. 
The positions hardly change as both the distance to reionization
and the distance between the two scattering surfaces
scales approximately in the same way with the matter density (in this 
calculations the matter density was given by $\Omega_0=1-\Omega_{\Lambda}$
where $\Omega_{\Lambda}$ is the energy density due to the cosmological
constant). 
On the other hand as the distance to
a fixed redshift increases with the cosmological constant, the 
optical depth $\kappa_{ri}$ increases, and consequently
the peaks should get higher. The fact that this is not the case
is a consequence of the COBE normalization, models with larger
values of the cosmological constant have larger additional
contributions to the low $l$ temperature anisotropies from the
ISW effect while polarization is not affected by the ISW.
Thus the changes in the normalization to keep the value of $C_{T10}$ fixed  
partially compensates the change in the height
of the new polarization peaks produced by the larger optical depth.   

Figure 3c and 3d explore the dependence of the polarization
power spectrum with the baryon density and the Hubble constant for
a fixed optical depth to decoupling, $\kappa_{ri}=1.0$. The rest of the
parameters where kept equal to those of standard CDM.
The height
of the first peak in the spectrum remains nearly constant as it is determined
by $\kappa_{ri}$ which was kept fixed. The fact that the peaks move is 
simple to understand, the redshift of reionization is given by
$(1+z_{ri})\approx 100 [\kappa_{ri} (0.5/h)(0.05/\Omega_b) (1/x_e)]^{2/3}$ 
and so $l$ scales
approximately as $l\propto (\kappa_{ri}/ h\Omega_b x_e)^{1/3}$. 

In the sCDM model reionization must have occurred extremely early
($z_{ri}\approx 100$)
in order to produce an optical depth of one;
even an optical depth of $\kappa_{ri}=0.5$ 
is only obtained for a redshift of $z_{ri}\approx 60$. But the
situation is different for open models or models with a cosmological
constant. An approximate scaling for the optical depth
valid for $\Omega_0 z_{ri}\gg 1$
is $\kappa_{ri}\propto (h \Omega_b x_e/\Omega_0^{1/2}) (1+z_{ri})^{3/2} $,
so for example reionization starting at $z_{ri}\approx 23$ will produce
an optical depth $\kappa_{ri}\approx 0.5$ in a model with
$\Omega_0=0.2$, $H_0=70 \rm km\ \rm sec^{-1}\ \rm 
Mpc^{-1}$ and $\Omega_b=0.1$.

\section{Measuring Polarization}

In this section I discuss the possibility of detecting 
polarization in the context of the standard theoretical models. 
I first concentrate  in an experiment like the one being 
built at Brown University and and in this case only in the detection of the
$rms$ degree of linear polarization and not on the measurement of the
correlation function. Then I  discuss the prospect of future satellite
missions like MAP.

\subsection{The Brown Experiment}

The Brown  experiment will try to measure 
both $Q$ and $U$ parameters with an expected sensitivity of  
$1 \mu \rm K$.  The instrument will allow measurements with a $7^{o}$
FWHM at an early stage and a $1^{o}$ FWHM afterwards. For concreteness
I will just take a gaussian window function, 
$W_l=\exp[-(l+0.5)^2 \sigma_{\theta}^2],\ \sigma_{\theta} = \theta
/ 2\sqrt{(2\ln2)}$ where $\theta$ is the FWHM of the detector in 
radians. The predicted values for the Stokes parameters were
calculated using CMBFAST and the spectra normalized to COBE. 
	        
First let us quote the expected {\it rms} value of $Q$ for standard 
CDM with no reionization, $P(7^{o})=4.8\times 10^{-2} \mu \rm K$ and 
$P(1^{o})=0.77 \mu \rm K$. These values, specially the large angular scale
one, are extremely small and thus very difficult to detect. This is the
reason why the reionized scenarios are the most promising to 
detect polarization.

Reionization will not only change 
the polarization power spectrum but also the temperature one, and
in some cases it may wash away the Doppler peaks completely. But
there is some degree of confusion between the different parameters
determining the CMB spectra, for example a reionization with a moderate
optical depth will decrease the amplitude of the Doppler peaks but 
this effect may be compensated by changing the spectral index
\cite{bond94}. In fact only an optical depth in the 
$10-20 \%$ range seem detectable from temperature maps alone
\cite{jung95}.
Figure 4 shows both polarization and temperature power 
spectra for standard CDM with a spectral index  $n=1$ and 
a reionized model with $\kappa_{ri}=0.5$ but a spectral index $n=1.2$. 
The difference in the anisotropy power spectrums is not so large,
while the polarization spectra are
 very different. The {\it rms} P values in 
this reionized case are
$P(7^{o})=1.2 \mu \rm K$ and $P(1^{o})=1.8 \mu \rm K$. 
For the large angular scale experiment the difference with 
standard CDM is more than two orders 
of magnitude and in the one degree case is more than a factor of two.
Thus a polarization measurement would easily distinguish
between the two scenarios.

Figure 5 shows the {\it rms} value of P as a function of $\kappa_{ri}$,
the major parameter determining  the amplitude of the 
polarization perturbation. $P(7^{o})$ only exceeds  $1 \mu \rm K$ level 
for $\kappa_{ri} \geq 0.5$ but saturates quickly near $1.8 \mu \rm K$.
On the other hand $P(1^{o})$ quickly raises above the $1\mu \rm K$ and
reaches  $3.2 \mu \rm K$ for an optical depth of two. This means that even
a negative detection at the  $1 \mu \rm K$ level for the one 
degree experiment 
is enough to rule out some models,
those with optical $\kappa_{ri} \ge 0.3$.

Parameters other than $\kappa_{ri}$ do not make much difference in the
height of the peaks. Table 1 explores the dependence
of $P(7^{o})$ and  $P(1^{o})$ with different cosmological parameters
for a fixed $\kappa_{ri}=1.0$. Although the the
height of the peaks remain almost constant in this models slight shifts 
in their  location  
change the predicted $P$. The $7^o$ $rms$ linear
polarization  is more sensitive to the position of the first peak. 
The $1^{o}$ experiment has the largest chance of 
putting interesting constrains on a possible reionization as the expected
signal is greater, because it is sensitive to all the power in the new
peaks of the polarization power spectrum.  A correlation
analysis between the polarization in the 
forty pixels that the experiment will measure
may help improve the above limits.


\subsection{Future Satellite Missions}

There are now two planned satellite misions to map the microwave
sky MAP \cite{map} and COBRAS/SAMBA \cite{cosa}
which will have polarization
information. Temperature information alone cannot put very stringent 
limits on the epoch of reionization \cite{jung95}.
With noise levels realistic for MAP only
$\kappa_{ri}\sim 0.1$ could be detected. The problem is that the 
dominant effect of reionization 
on the temperature on small angular scales is a 
suppression equivalent to a decrease in the amplitude of the
primordial perturbations. This degeneracy is broken on large
scales as reionization does not significantly affect
the amplitude on these scales, 
but here cosmic variance precludes very accurate 
determinations. One may hope to improve the accuracy in the
estimation of the optical depth by measuring the new 
peaks in the polarization power spectra.   

Figure 6 illustrates this points. In panel a. the spectra for
a COBE normalized sCDM and a reionized model with $\kappa_{ri}=0.1$
are plotted. The reionized model has been normalized in such a way
as to minimize the $\chi^2$ difference between the two.
I have assumed for simplicity that each $C_l$ is gaussian distributed
with a variance given by \cite{jung95}
\begin{equation}
\sigma_l=\sqrt{2\over 2l+1}[C_l+w^{-1}\exp(l^2\sigma_b^2)]
\end{equation}
where $\sigma_b^2=7.42\times 10^{-3}(\theta_{fwhm}/1^o)$ for
a gaussian beam and $w^{-1}$ is a pixel size independent measure
of experimental noise. Values corresponding to the MAP mission
where used ($w^{-1}=4.2\times 10^{-15}$ and $\theta_{fwhm}=0.29^o$).

Figure 6b shows the polarization power spectra for the same models,
the difference in the large scale polarization greatly exceeds the
cosmic variance. The value of the multipoles $C_{El}$
at the reionization peak in this model are $C_{El}\sim (0.12 \mu \rm K)^2$,
to be compared to a noise in each $a_{lm}$ of roughly $0.14 \mu \rm K$
for polarization \cite{sper96} in the case of the MAP mission. 
This makes
the possibility of using polarization to further constrain the optical
depth very interesting. It is also
worth noting that the noise levels of COBRAS/SAMBA detectors is much lower,
and so better sensitivities should be expected in this case.

\section{Conclusions}

The polarization of the microwave background is very sensitive to the 
ionization history of the universe and an early reionization can 
greatly enhance it.
I have discussed in detail the physics behind the generation
of polarization in reionized scenarios and the appearance of new peaks in
the polarization power spectrum. I have identified the major parameters
determining the location of these peaks, the ratio of distances between 
the observer and the reionization scattering surface to that between 
reionization and recombination. The height of the peaks is mainly function of
$\kappa_{ri}$, the optical depth to recombination.

An early reionization with an optical depth $\kappa_{ri}\geq 0.5$
can take large and intermediate angular scale 
polarization to the $\mu \rm K$ level, detectable in the near future 
by the Brown Experiment.
Polarization may help resolve some of the
``confusion'' that can arise when determining 
cosmological parameters using CMB. In particular it may help
detect levels of reionization below the $\kappa_{ri}\sim 0.1$
that can be obtained with temperature maps alone.

\section*{Acknowledgments}
I am very grateful to Uro\v s Seljak for his encouragement 
and many useful discussions. I would also want to thank
Ed Bertschinger and David Spergel for reading the manuscript.

\newpage

\vspace {10cm}

\begin{tabular}{||l|l|l|l|l||}    \hline
$\Omega_0$&$\Omega_b$&$H_0$&$P(7^o)$&$P(1^o)$ \\ \hline
1.0 & 0.05 & 50 & $4.81\, 10^{-2}$& 0.642 \\ \hline
0.7 & 0.05 & 50 & 1.62 & 2.25 \\ \hline
0.5 & 0.05 & 50 & 1.67 & 2.50 \\ \hline
0.3 & 0.05 & 50 & 1.62 & 2.25 \\ \hline
1.0 & 0.03 & 50 & 1.40 & 2.67 \\ \hline
1.0 & 0.08 & 50 & 1.83 & 2.79 \\ \hline
1.0 & 0.10 & 50 & 1.91 & 2.80 \\ \hline
1.0 & 0.05 & 60 & 1.72 & 2.79 \\ \hline
1.0 & 0.05 & 80 & 1.84 & 2.85 \\ \hline
1.0 & 0.05 &100 & 1.92 & 2.88 \\ \hline
\end{tabular}

\vspace{5cm}
\noindent{
Table 1.
Degree of linear polarization in $\mu \rm K$ 
SCDM (first row) and several other models
all with $\kappa_{ri}=1.0$. The value of the
cosmological constant is such that all the above 
models are flat, $\Omega_{total}=1.0$. $H_0$ is the Hubble constant in 
$\rm km \; \rm sec^{-1} \rm Mpc^{-1}$.}

\newpage

\begin{figure}[t]
\centerline{\psfig{figure=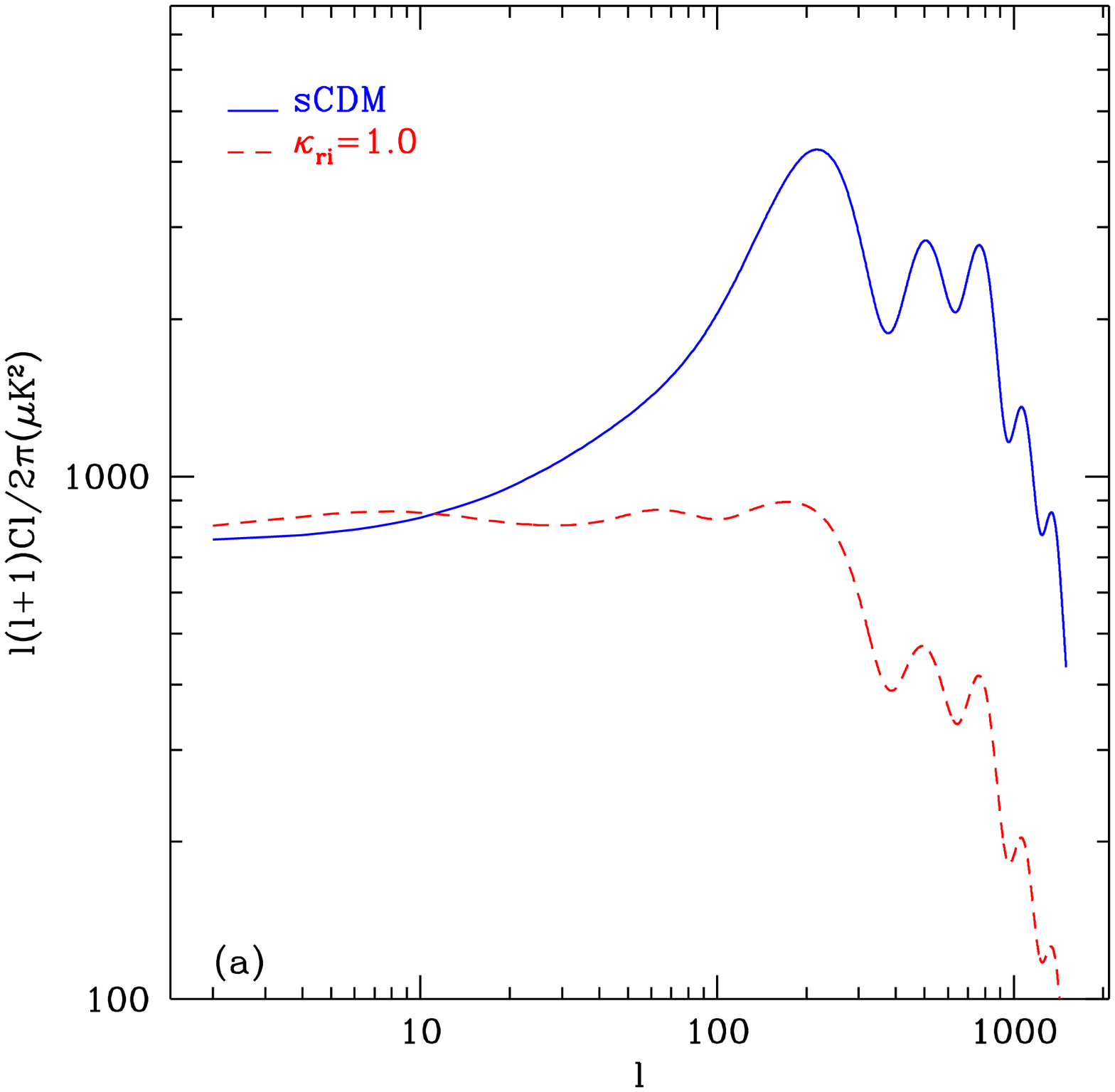,height=3.7in,width=5in}}
\centerline{\psfig{figure=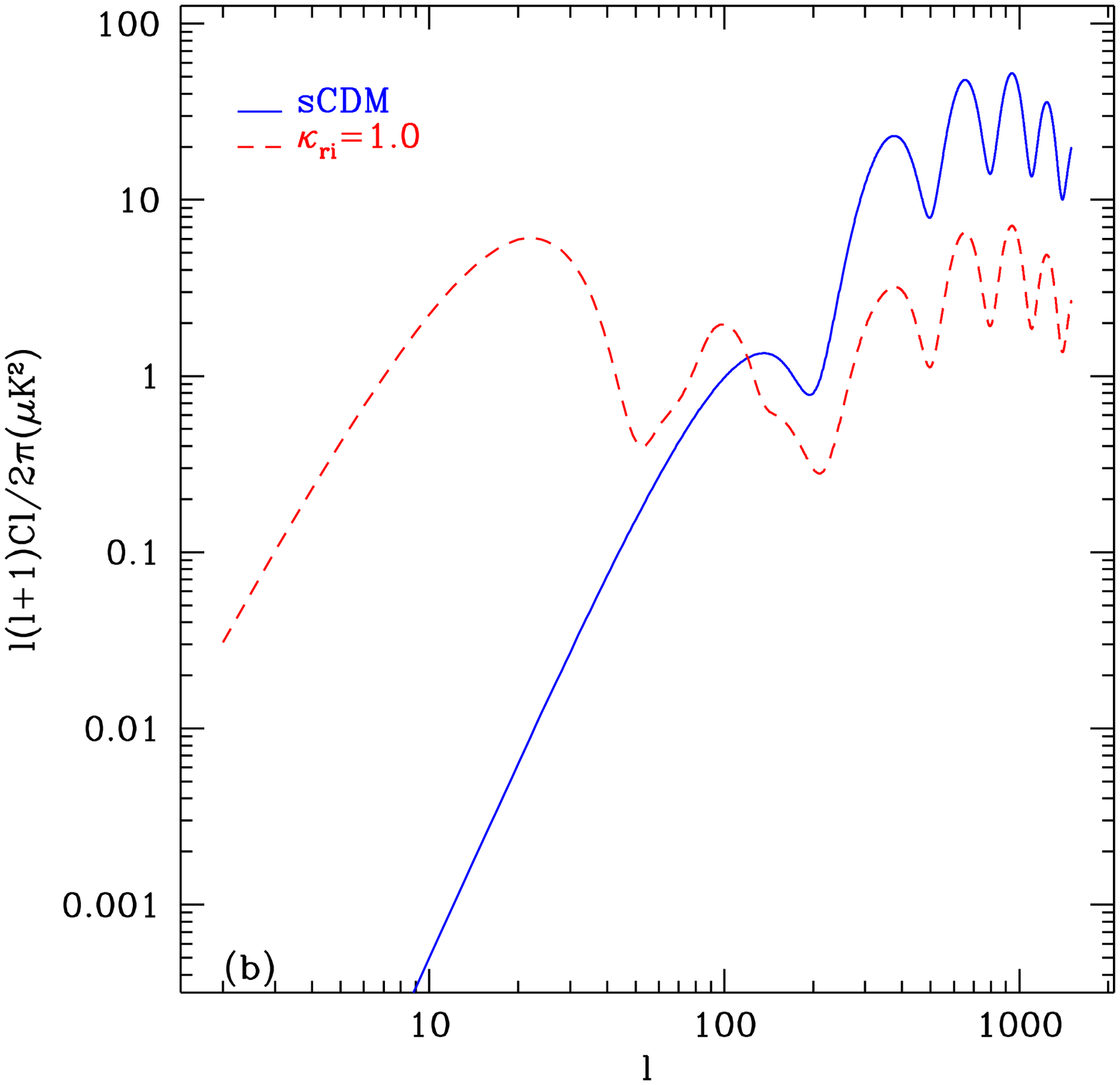,height=3.7in,width=5in}}
\caption{$l(l+1)C_{l}  / 2\pi$ for both temperature (a) and
polarization (b) for standard CDM and a model where the optical depth to 
recombination is $\kappa_{ri}=1.0$.}
\label{fig1}
\end{figure}

\newpage

\begin{figure}[t]
\centerline{\psfig{figure=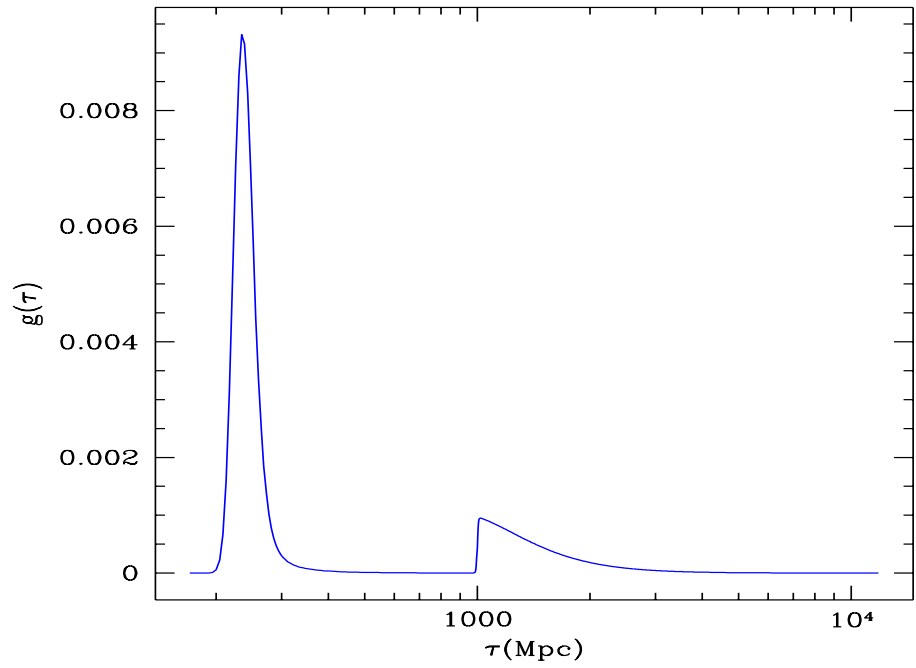,height=3.7in,width=5in}}
\label{fig2}
\caption{Visibility function for standard CDM with reionization
such that the optical depth to 
recombination is $\kappa_{ri}=1.0$.} 
\end{figure}

\newpage

\begin{figure}[t]
\centerline{\psfig{figure=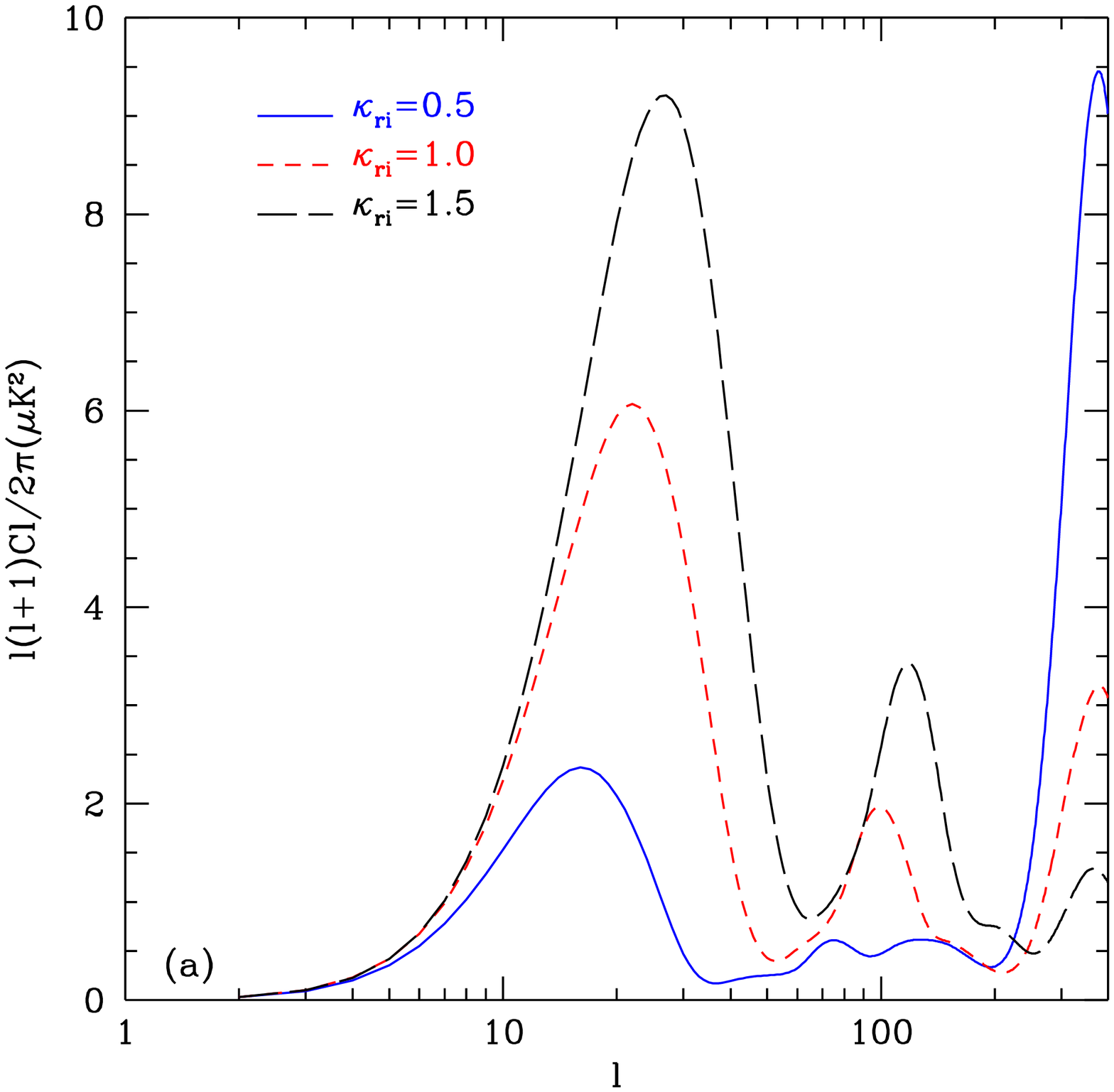,height=3.5in,width=5in}}
\centerline{\psfig{figure=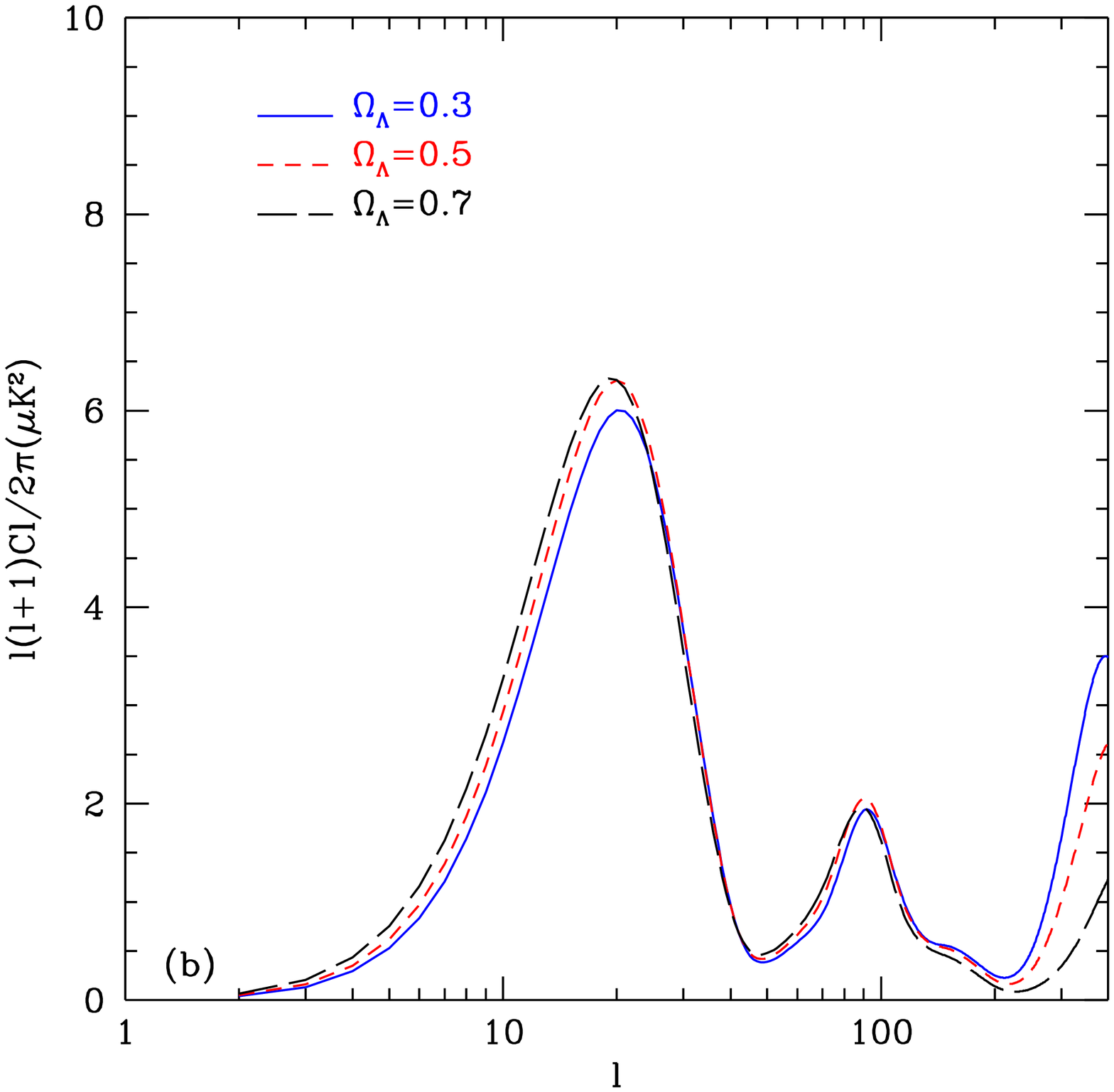,height=3.5in,width=5in}}
\end{figure}

\newpage

\begin{figure}[t]
\centerline{\psfig{figure=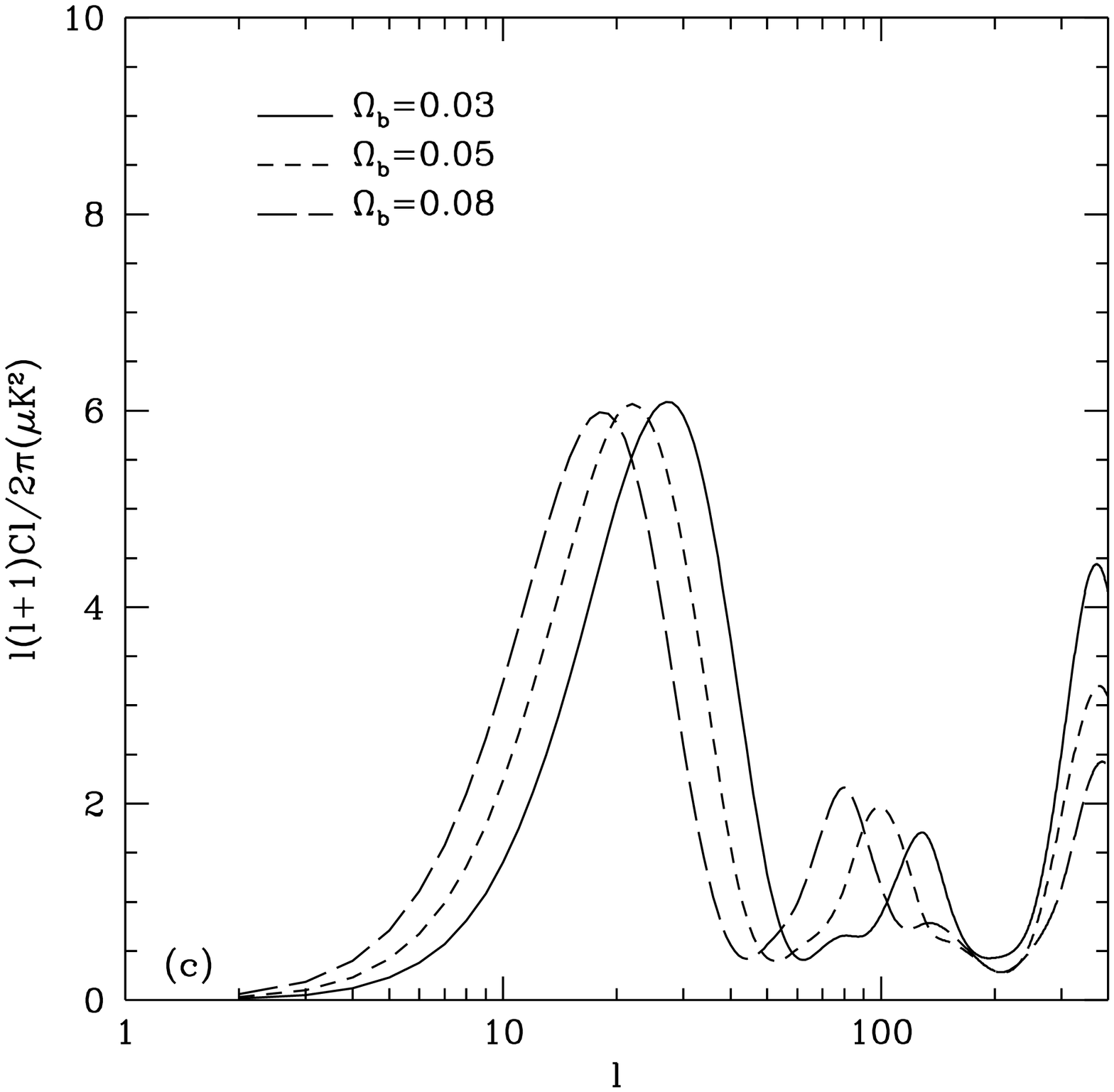,height=3.5in,width=5in}}
\centerline{\psfig{figure=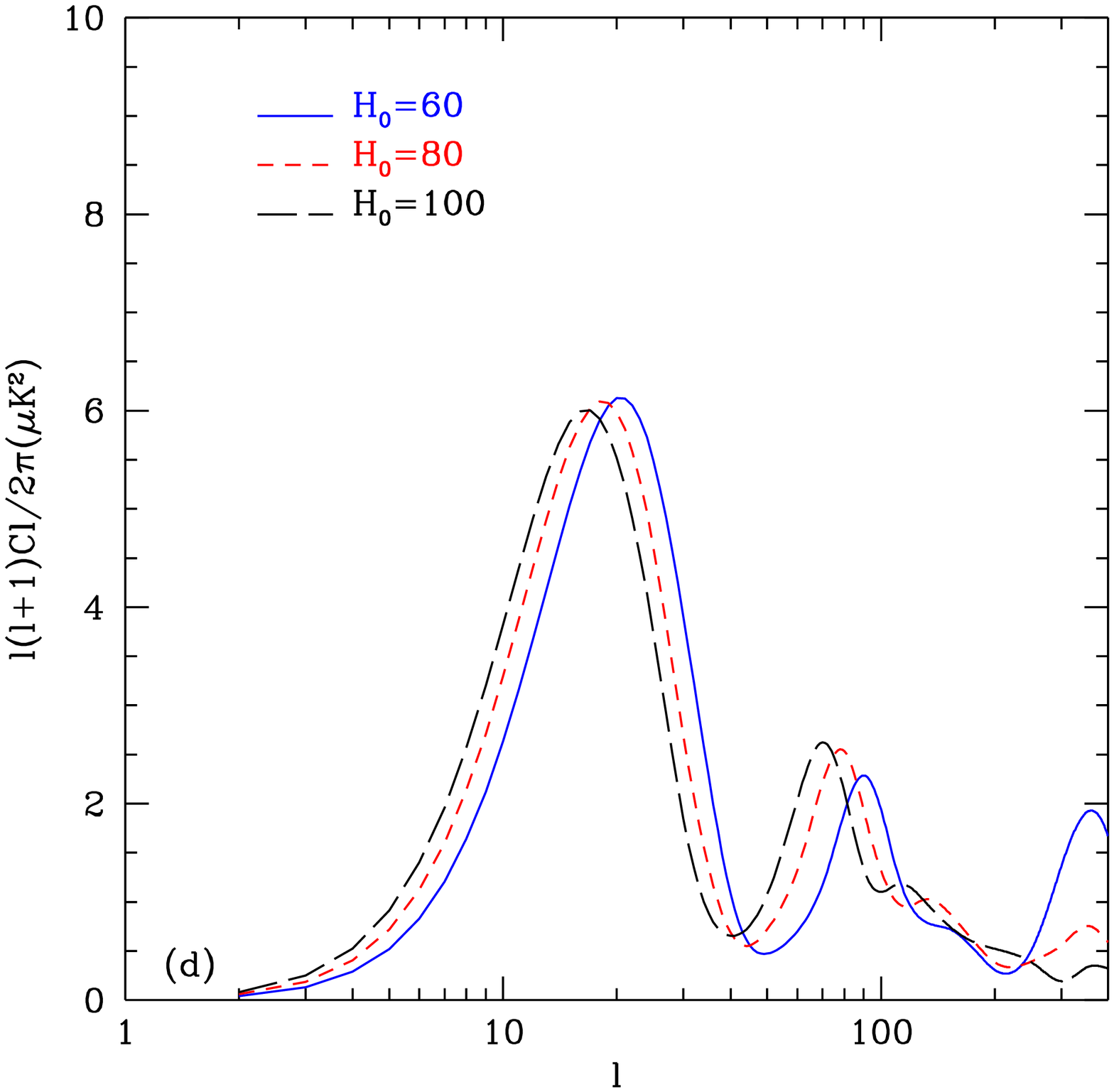,height=3.5in,width=5in}}
\caption{$l(l+1)C_{lP} / 2\pi$ (a) for CDM models with
varying  $\kappa_{ri}=0.5,\ 1.0,\ 1.5$ and 
(b) for models with varying cosmological constant $\Omega_{\Lambda}
=0.3,\ 0.5,\ 0.7$ and a fixed redshift of reionization $z_{ri}=100$. 
Reionized ($\kappa_{ri}=1.0$) CDM models (c) with
varying  $\Omega_b=0.3,\ 0.5,\ 0.8$ and 
(d) with different Hubble costants $Ho
=60,\ 80,\ 100 \ \rm km \; \rm sec^{-1} \rm Mpc^{-1}$. In all cases
reionization was assumed to be total ($x_e=1$)}
\label{fig3}
\end{figure}

\newpage

\begin{figure}[t]
\centerline{\psfig{figure=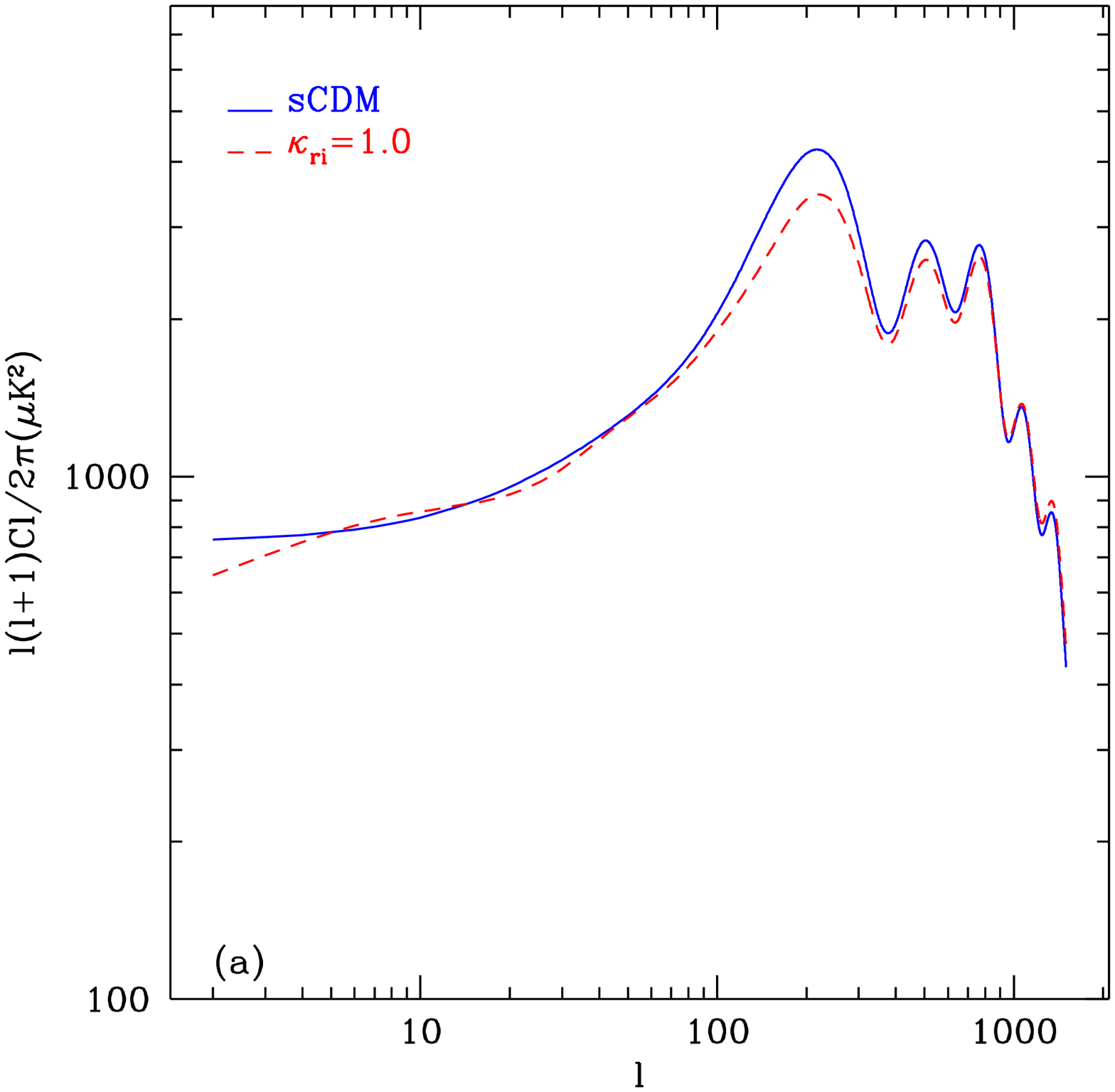,height=3.7in,width=5in}}
\centerline{\psfig{figure=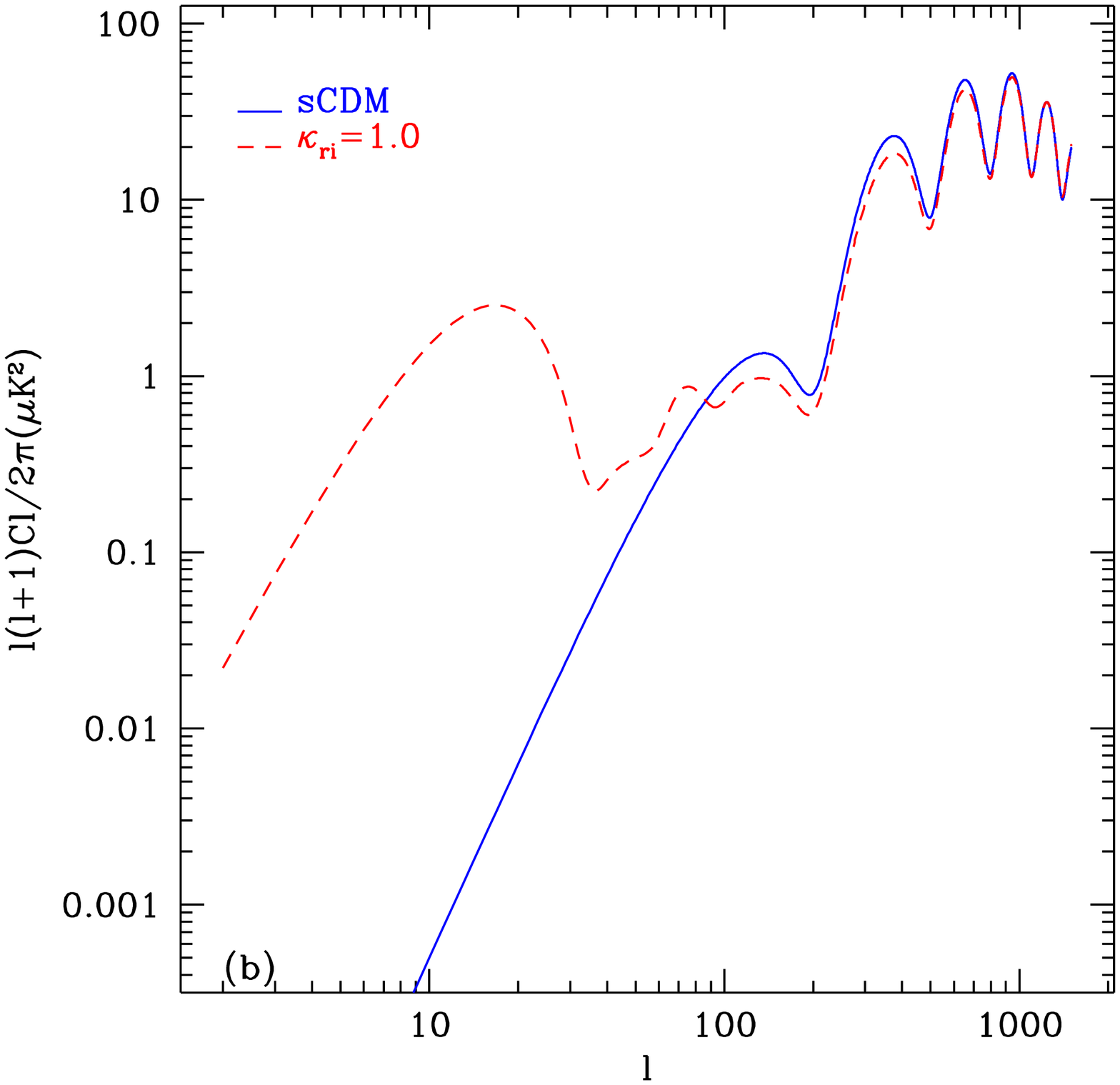,height=3.7in,width=5in}}
\caption{$l(l+1)C_{l} / 2\pi$ for both temperature (a) and
polarization (b) for standard CDM and a model where the optical depth to 
recombination is $\kappa_{ri}=0.5$ and a spectral index $n=1.2$.}
\label{fig4}
\end{figure}

\newpage

\begin{figure}[t]
\centerline{\psfig{figure=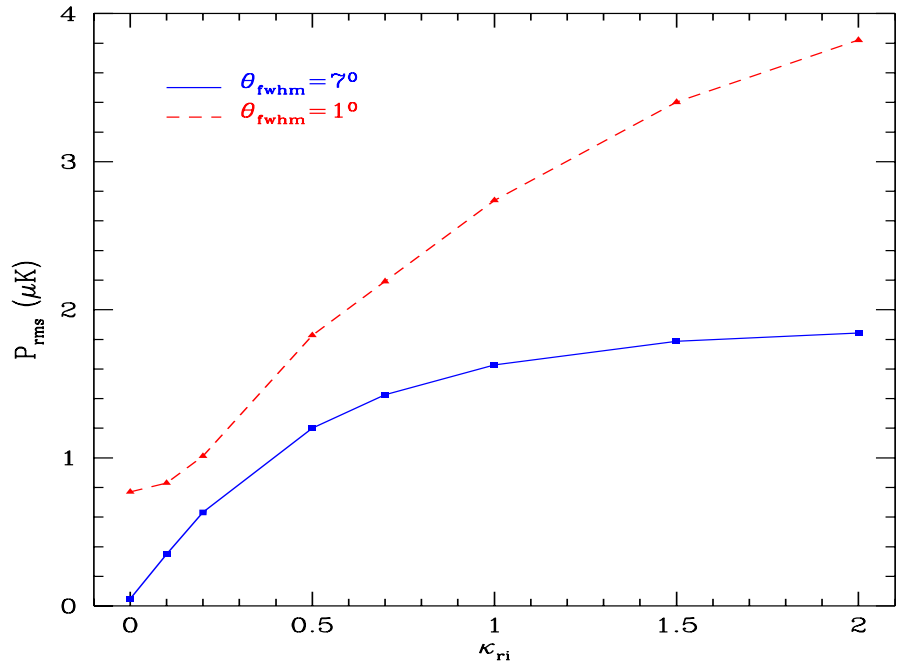,height=3.7in,width=5in}}
\caption{Polarization {\it rms} fluctuations ($\mu K$) as a function of 
the optical depth, $\kappa_{ri}$ for a $7^o$ and $1^o$ FWHM experiments.}
\label{fig5}
\end{figure}
\newpage

\begin{figure}[t]
\centerline{\psfig{figure=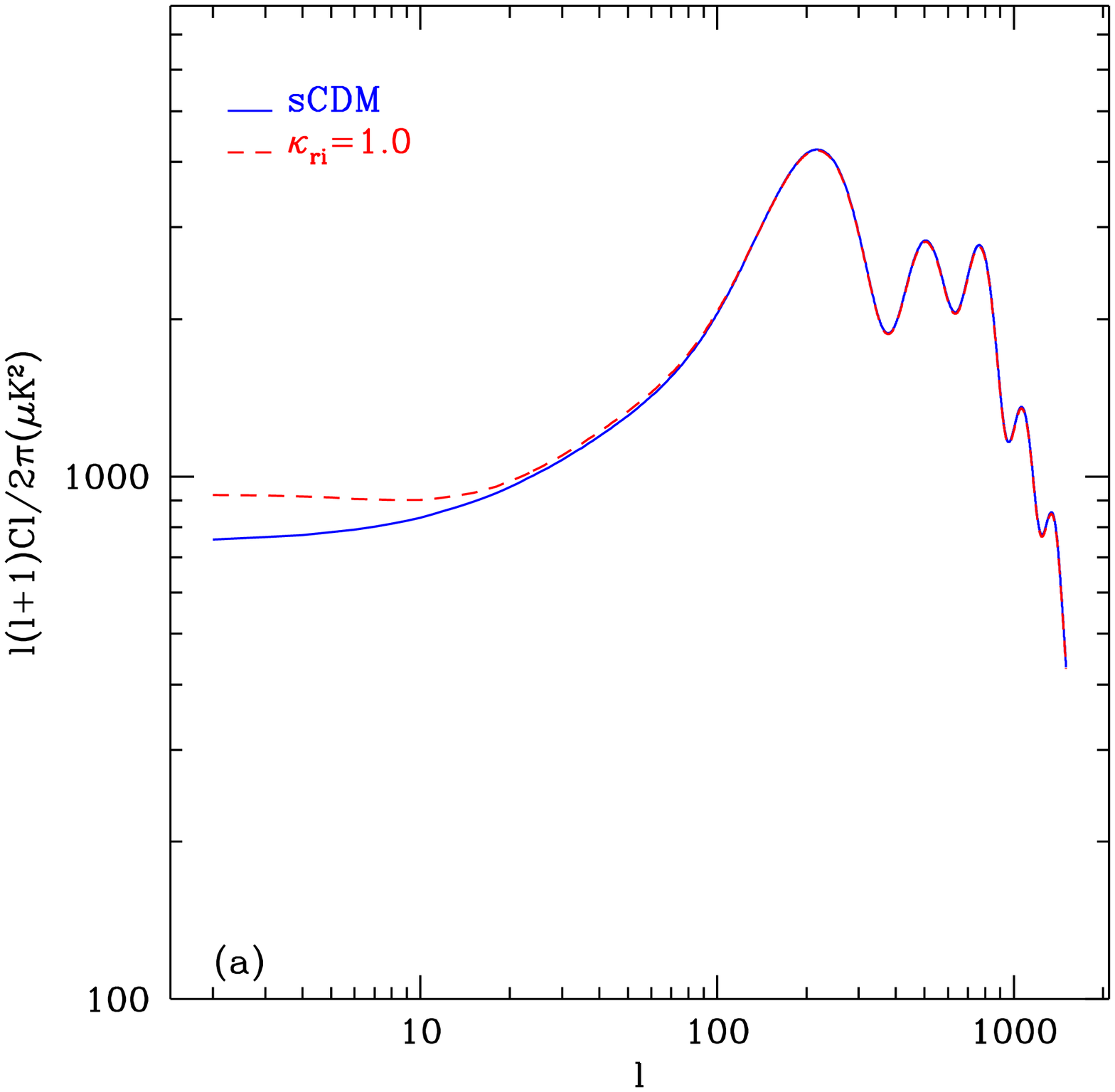,height=3.7in,width=5in}}
\centerline{\psfig{figure=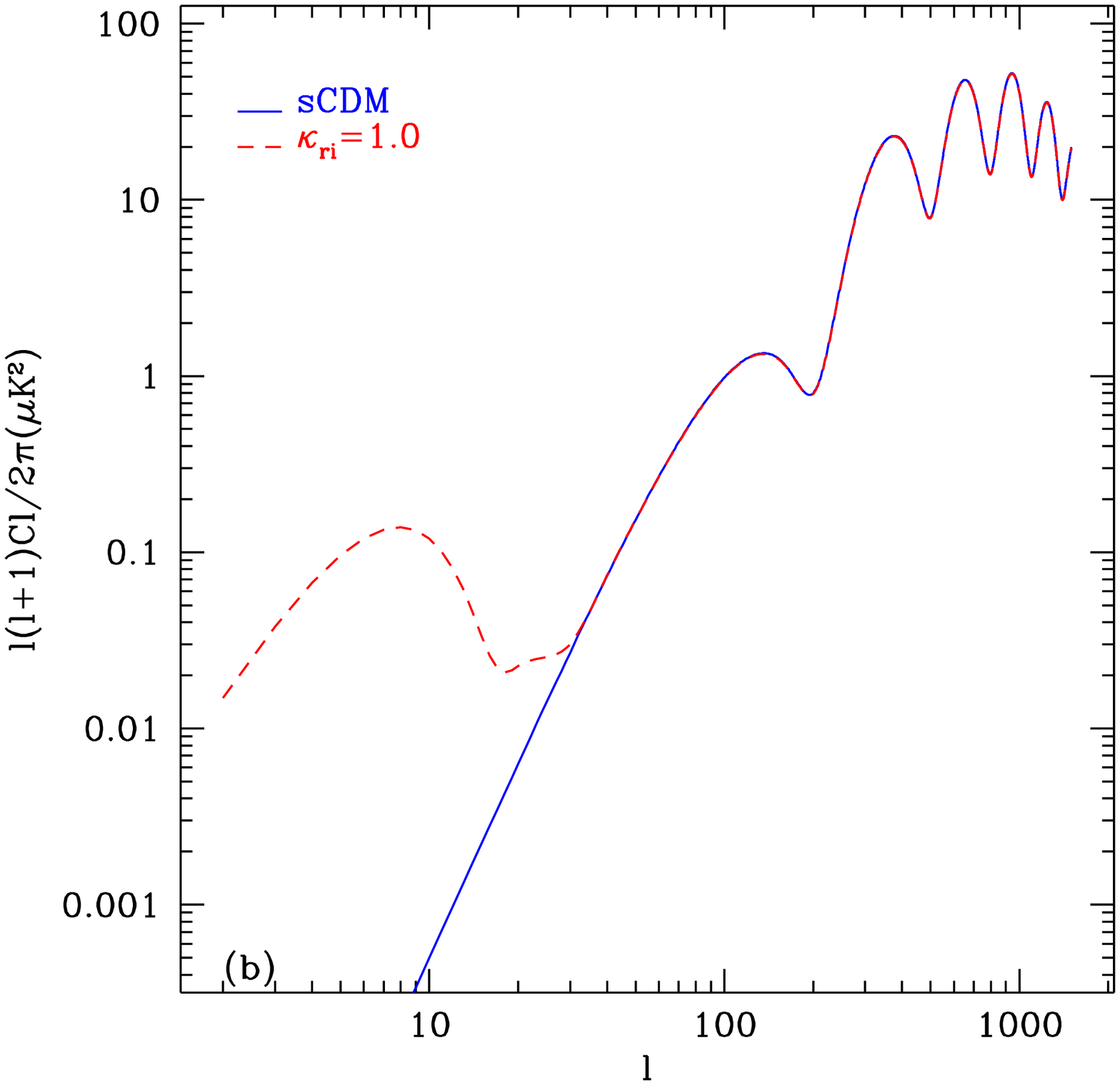,height=3.7in,width=5in}}
\caption{Temperature and Polarization 
power spectra for a COBE normalized sCDM
and a reionized model with $\kappa_{ri}=0.1$.}
\label{fig6}
\end{figure}

\end{document}